\documentclass[a4paper,11pt]{article}
\pdfoutput=1
\usepackage{jcappub}
\usepackage[T1]{fontenc}
\usepackage{hyperref}
\usepackage{graphics}
\usepackage{amssymb, amsmath, natbib}
\usepackage[dvipsnames]{xcolor}
\usepackage{hhline}
\usepackage{multirow}
\usepackage{placeins}
\usepackage{afterpage}
\usepackage{float}

\newcommand{\br}{\mathbf{r}}
\newcommand{\bs}{\mathbf{s}}

\newcommand{\bx}{\mathbf{x}}
\newcommand{\bk}{\mathbf{k}}

\newcommand{\rsp}[1]{{\color{black}#1}}

\title{Mind the gap: the power of combining photometric surveys with intensity mapping}

\author[a,b]{Chirag Modi,}
\author[a, c]{Martin White,}
\author[d,e]{Emanuele Castorina,}
\author[e]{An\v{z}e Slosar}

\affiliation[a]{Berkeley Center for Cosmological Physics, Department of Physics, University of California, Berkeley, CA 94720}
\affiliation[b]{Center for Computational Astrophysics, Flatiron Institute, 162 Fifth Ave., New York, NY 10010, USA}
\affiliation[c]{Department of Astronomy, University of California, Berkeley, CA 94720}
\affiliation[d]{Dipartimento di Fisica ‘Aldo Pontremoli’, Universit\`a degli Studi di Milano, Milan, Italy}
\affiliation[e]{Theoretical Physics Department, CERN, 1211 Geneva 23, Switzerland}
\affiliation[f]{Department of Physics, Brookhaven National Laboratory, Upton, NY 11973}

\emailAdd{cmodi@flatironinstitute.org}
\emailAdd{mwhite@berkeley.edu}
\emailAdd{emanuele.castorina@unimi.it}
\emailAdd{anze@bnl.gov}

\keywords{cosmological parameters from LSS -- 21 cm -- galaxy clustering --bias model -- forward modeling}

\abstract{The long wavelength modes lost to bright foregrounds in the interferometric 21-cm surveys can partially be recovered using a forward modeling approach that exploits the non-linear coupling between small and large scales induced by gravitational evolution.
In this work, we build upon this approach by considering how adding external galaxy distribution data can help to fill in these modes.
We consider supplementing the 21-cm data at two different redshifts with a spectroscopic sample (good radial resolution but low number density) loosely modeled on DESI-ELG at $z=1$ and a photometric sample (high number density but poor radial resolution) similar to LSST sample at $z=1$ and $z=4$ respectively.
We find that both the galaxy samples are able to reconstruct the largest modes better than only using 21-cm data, with the spectroscopic sample performing significantly better than the photometric sample despite much lower number density.
We demonstrate the synergies between surveys by showing that the primordial initial density field is reconstructed better with the combination of surveys than using either of them individually.
Methodologically, we also explore the importance of smoothing the density field when using bias models to forward model these tracers for reconstruction.
}

\begin{document}
\maketitle
\flushbottom

\section{Introduction}
\label{sec:intro}

The study of large-scale structure in the high-redshift Universe is a promising tool for cosmology \cite{Ferraro19}.  One means of mapping large-scale structure in the distant Universe is through the technique of intensity mapping (IM): performing a low resolution, spectroscopic survey to measure integrated flux from unresolved sources on large areas of sky at different frequencies. Such surveys capture the largest elements of the cosmic web and map out the distribution of matter in very large cosmological volumes in a fast and efficient manner, with good radial resolution \cite{Kovetz17,Kovetz19}.  Since hydrogen is so abundant in the Universe, 21-cm emission from cosmic neutral hydrogen (H{\sc i}) offers one tracer to map out the Universe in such a way.  With its low energy and optical depth there is little chance of line confusions and it provides an efficient way to probe the spatial distribution of neutral hydrogen, and hence the underlying dark matter from the local Universe to the dark ages \cite{Furlanetto06,SKACosmo,SKAIM,Kovetz19,CVDE-21cm,PUMARFI}.

One issue with IM surveys is that foregrounds render the long-wavelength fluctuations along the line of sight unmeasureable, which can adversely affect the science that they can do \cite{Seo16,Cohn16,Obuljen2017,Chen19,Modi19b}.  In ref.~\cite{Modi19b} we studied one method for reconstructing long-wavelength fluctuations, using the distinct pattern of correlations imprinted by gravitational instability.  In this paper we take another route, more similar to refs.~\cite{Cohn16,Chen19}, and consider how adding additional data can help to fill in the modes that are lost to foregrounds in 21-cm observations.

\begin{figure}
    \centering
    \resizebox{\columnwidth}{!}{\includegraphics{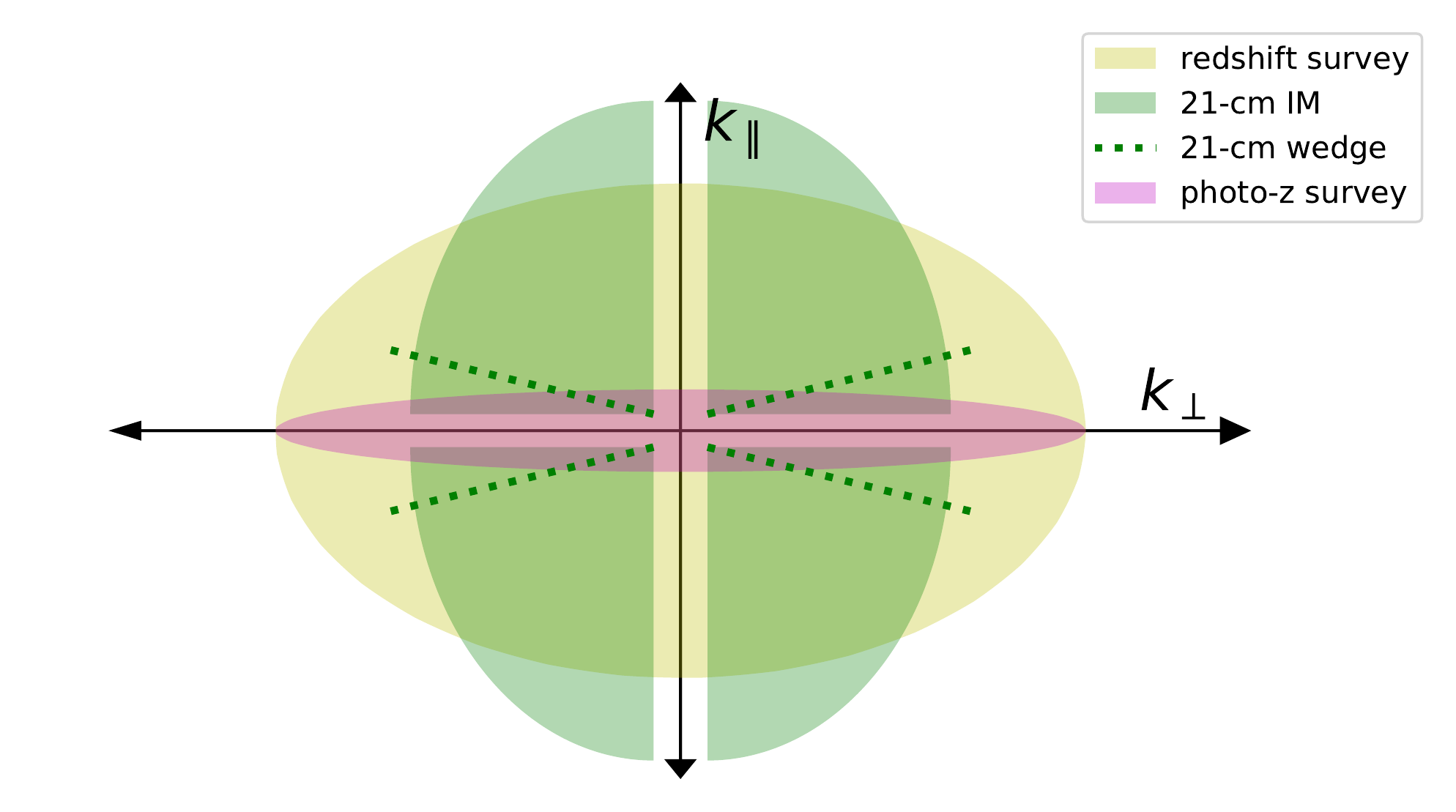}}
    \caption{A schematic view of the modes probed by 21-cm surveys and optical surveys in the $k_\perp-k_\parallel$ plane.  Photometric redshift surveys (purple) are capable of high angular resolution (i.e.~probe to high $k_\perp$) but have limited radial resolution and thus only constrain well low $k_\parallel$ modes.  A spectroscopic redshift survey (yellow) would enable access to the full $\vec{k}$ plane, but would be prohibitively expensive if done with high number density.  An IM survey of 21-cm emission (green) has high number density and radial resolution but limited angular resolution and misses modes at low $k_\parallel$ (and in a region known as the foreground wedge).}
    \label{fig:schematic}
\end{figure}

\section{Mock samples}

To model our galaxy and IM surveys we use an extension of the Hidden Valley simulations \cite{Modi19a}, a set of $10240^3$ particle N-body simulations performed in a $1024\,h^{-1}$Mpc box using the {\sc FastPM} code \cite{FengChuEtAl16}.  Further details of this simulation and the manner in which {\sc Hi} is assigned to halos can be found in Appendix \ref{app:HI}. We consider the data in the redshift space. Anticipating that the {\sc Hi} would be observed by an instrument such as the Hydrogen Intensity and Real-time Analysis eXperiment (HIRAX; \cite{HIRAX}) or the Packed Ultrawideband Mapping Array (PUMA \cite{PUMA,PUMARFI}) we assign the {\sc Hi} to a regular $512^3$ grid and work with the Fourier transform of the density field.  To the signal we add thermal noise and foregrounds as described in detail in refs.~\cite{Modi19a,Modi19b}. For $z=1$ we will present results for both PUMA and HIRAX thermal noise, while for $z=4$ we will have observations only from PUMA and so restrict ourselves to that case.

Due to foregrounds and difficulties in instrument calibration, we assume the {\sc Hi} map has infinite noise for modes with $k_\parallel\approx 0$ and in a region of the $k_\perp-k_\parallel$ plane known as the ``wedge'' (see Fig.~\ref{fig:schematic}).  For details of how these limitations arise, we refer the reader to refs.~\cite{Modi19a,Modi19b} and the extensive references to the earlier literature.  Our goal here is to ask to what extent a different survey, either a photometric or spectroscopic galaxy survey, can ``fill in'' these missing modes.  Ideally the galaxy survey can take advantage of the high number density and excellent radial resolution of the {\sc Hi} survey while the {\sc Hi} survey can take advantage of the low-$k_\parallel$ sensitivity of the galaxy survey. We will present results for a `pessimistic' choice of foreground wedge which removes information for angles less than three times the primary field of view ($3\times$FOV). This corresponds to a wedge angle of $\theta_w = 6^\circ$ and $38^\circ$ at $z=1$ and $z=4$ respectively \cite{Modi19b}. \rsp{In our notation, this means we exclude all modes with 
\begin{align}
    k_{||} < \sin(\theta_{w})\frac{D(z)H(z)}{1+z}k_\perp\,.
\end{align}
where $D(z)$ is the line-of-sight comoving distance.
We also remove all modes with $k_{||}<0.1\,\text{Mpc}/h$.}
Obviously a better instrument calibration and foreground subtraction, which leads to a smaller wedge, will require less input from the auxilliary data.

We will consider two populations of mock galaxies, loosely modeled on samples that might be returned by upcoming surveys. Both the galaxy samples will again be considered in the redshift space.
At $z=1$ we consider a galaxy sample with good redshift measurements and $\bar{n}=10^{-3}\,h^3\,{\rm Mpc}^{-3}$.  Such a sample is similar to the emission line galaxy (ELG) sample to be targeted by the Dark Energy Spectroscopic Instrument (DESI) survey \cite{DESI}, and henceforth we will refer it so.  For this exploratory calculation, rather than model these galaxies in great detail, we simply choose a mass-limited sample of halos in our N-body simulation with $\bar{n}=10^{-3}\,h^3\,{\rm Mpc}^{-3}$.  We assume the redshifts of these halos are known precisely.  Such a halo sample has a complex, scale-dependent bias and about the right level of shot noise while being very easy to model.  We do not expect our results to depend upon the details of this choice.

The second sample, appropriate for $z\geq 1$, is a photometric sample of galaxies such as will be observed by the Vera Rubin Observatory - Legacy Survey of Space and Time (LSST; \cite{LSST}) and thus we will refer to it as LSST sample.  We follow ref.~\cite{Wilson19} and consider an analogue of Lyman break dropout galaxies (LBGs), \rsp{which is fairly typical of proposed future surveys \citep{Modi17, Ferraro19, Schlegel19}}.  This sample has high number density but poor radial resolution, leading to a density field with lower noise at low $k_\parallel$ but high noise for large $k_\parallel$. 
As above we model these galaxies as a mass-limited halo sample.  We introduce the photometric redshift scatter simply by enhancing the  noise for this sample as an exponential in $k_\parallel$ (see \ref{eq:photosmoothing}). 
We will consider the LSST sample at two redshifts, $z=1$ and $z=4$ with number densities $\bar{n}=5 \times 10^{-2}\,h^3\,{\rm Mpc}^{-3}$ and $\bar{n}=3.5 \times 10^{-3}\,h^3\,{\rm Mpc}^{-3}$ respectively.
\rsp{We expect our results to qualitatively remain the same for other photometric samples as long as they can be described with a bias model and redshift scatter,
even if the specific gains may vary depending on the number density and photometric smoothing.}

\section{Method}

Our reconstruction method largely follows the steps outlined in Ref.~\cite{Modi19b}. We reconstruct the initial density field by optimizing its posterior, conditioned on the observed data ({\sc Hi} and galaxy density in different regions of $k$-space), assuming Gaussian initial conditions. Evolving this initial field allows us to reconstruct the observed data on all scales. The initial conditions are reconstructed in the manner described in refs.~\cite{Seljak17,Modi18,Modi19b} (for alternative approaches to reconstruction see also refs.~\cite{Jasche, Wang14,Zhu:2016sjc, Karacay19, Schmittfull17}).

For the forward model [$\mathcal{F}(\bs)$] connecting the observed data $(\mathbf{\delta})$ with the Gaussian initial conditions (ICs; $\mathbf{s}$) we use a second order Lagrangian bias model coupled to the non-linear dynamics of the simulation.
\rsp{We explore different non-linear dynamics for the gravitational evolution and find the best results when using Zeldovich dynamics to evolve particles from their Lagrangian to Eulerian position.}
The forward-modeled final density fields (of galaxies and {\sc Hi}) are obtained by assigning particles to a grid at their final redshift-space positions with a weight that is a function of the density and shear at their positions in the initial conditions \cite{Modi19b}.  
Our Lagrangian bias model \cite{Matsubara08a,Matsubara08b,Carlson13,White14,Vlah16,Modi16,Schmittfull19,Modi19b,Modi20} connects the matter field to the dark matter halos and includes terms up to quadratic order ($\delta_L$, $\delta_{L,R}^2$ and $s^2_{L,R} \equiv \sum_{ij}s_{ij}^2$  the scalar shear field\footnote{Since $\delta_L^2$ and $s^2$ are correlated, we actually define a new field $g_L^2 = \delta_L^2- s^2$ which does not correlate with $\delta_L^2$ on large scales and use this instead of shear field.} where $s_{ij}^2 = (\partial_i \partial_j\partial^{-2} - [1/3]\delta_{ij}^{D}) \delta_L$) computed from the ICs of the simulation, evolved to $z=0$ using linear theory. 
We subtract the zero-lag terms to make these fields have zero mean.

To construct the bias terms of our model, and in particular to estimate the quadratic operators, we smooth the linear density field with a Gaussian kernel with smoothing scale $R$. 
The dependence on the smoothing scale $R$ of reconstruction algorithms is an open problem in the field \cite{Schmidt:2018bkr,Cabass:2020nwf}, since a formal understanding of the renormalization of the bias expansion at the field level has not been obtained yet \cite{Schmittfull19,Cabass:2020nwf}. It should also be kept in mind that, even without any additional smoothing, the size of the FFT grid provides an unavoidable cut off of the power on small scales.
We thus explore different smoothing scales for reconstruction and empirically motivate our choices. 
For $z=4$, any smoothing leads to lower reconstruction performance, implying that the optimal smoothing is likely smaller than the grid resolution. 
For $z=1$, we find that when reconstructing with only {\sc Hi} data, in which case the only large scale information is provided by non-linear coupling of gravitational evolution, the large scales are best reconstructed when smoothing the \rsp{linear field in the forward model of {\sc Hi} data} with $R=6\,h^{-1}$Mpc.
However when combining this with additional data of galaxy field, we find that no smoothing is required since the information on large scales is dominated by the galaxy field. 
Thus for $z=1$, we smooth \rsp{the linear field in the forward model of {\sc Hi} data} with $R=6\,h^{-1}$Mpc
and do not smooth the linear field for the galaxy data. 
We intend to the return to the issue of the smoothing scale in a forthcoming publication.

Our modeled tracer field is then \cite{Modi19b,Modi20,Kokron21}:
\begin{equation}
    \delta_{\rm HI}^b(\bx) = \delta_{[1]}(\bx) + b_1\delta_{[\delta_L]}(\bx) + b_2\delta_{[\delta^2_{L, R}]}(\bx)  + b_g \delta_{[g_{L, R}]}(\bx) \qquad .
\label{eq:biasmodel1}
\end{equation}
where $\delta_{[W]}(\bx)$ refers to the field generated by weighting the particles with the `$W$' field.

To fit\footnote{In principle one could fit for the bias parameters using summary statistics, such as the power spectrum, as in e.g.\ refs.\ \cite{Modi20,Kokron21}, though we anticipate that the constraints from the field itself would be tighter.} for the bias parameters, we minimize the mean square model error between the data and the model fields which is equivalent to minimizing the \rsp{error power spectrum i.e.\ the power spectrum of the residuals between the bias model and true (clean) data, $\br(k) = \delta^b(\bk) - \delta^{\mathrm data}(\bk)$, in Fourier space}.  We do this separately for the {\sc Hi} and galaxy fields and thus have two sets of bias parameters.
The smoothing scale affects the forward modeling and reconstruction differently. The accuracy of forward modeling the observation from true initial conditions is not as sensitive but the accuracy of the reconstructed field from observed data gets impacted more significantly.
Thus ideally one would like to keep both the bias parameters and the smoothing scale as a free-parameters to be fit at the time of reconstruction instead of fitting them in advance.
We plan to explore this in the future.

Once the bias parameters are known, we reconstruct the initial (and final density) field by maximizing the posterior as a function of the IC amplitudes, $\mathbf{s}(\mathbf{k})$, using L-BFGS\footnote{https://en.wikipedia.org/wiki/Limited-memory\_BFGS} \cite{nocedal06}.  The negative log-likelihood for the Gaussian prior can be combined with the negative log-likelihood of the data to get the posterior (see also \cite{Schmittfull19,Schmidt:2018bkr})
\begin{align}
    \mathcal{P} &=  \sum_k \frac{1}{\rm N_{modes}(k)} \left( \sum_{\substack{\bk, |\bk|\sim k, \\ {\bk \not\in w}}}  \frac{|\delta_{\rm HI}^b(\bk) - \delta_{\rm HI}^{\rm obs}(\bk)|^2}{P_{\rm err-HI}(k,\mu)} \right)
    \nonumber \\
    % &+ \sum_k \frac{1}{\rm N_{modes}(k)} \left( \sum_{\bk, |\bk|\sim k}  \frac{|\delta_{\rm g}^b(\bk) - \delta_{\rm g}^{\rm obs}(\bk)|^2}{P_{\rm sn}(k,\mu)}
    &+ \sum_k \frac{1}{\rm N_{modes}(k)} \left( \sum_{\bk, |\bk|\sim k}  \frac{|\delta_{\rm g}^b(\bk) - \delta_{\rm g}^{\rm obs}(\bk)|^2}{P_{\rm err-g}(k)}\right) \nonumber \\
    &+ \sum_k \frac{1}{\rm N_{modes}(k)} \left(  \sum_{\bk, |\bk|\sim k}  \frac{|\bs(\bk))|^2}{P_{\rm s}(k)}  \right)
\label{eq:posterior}
\end{align}
where $P_{\rm s}$ is the prior power spectrum of the initial conditions and the sum is over modes that are measured by each survey (i.e.~modes not in the foreground wedge for {\sc Hi} and low $k_\parallel$ modes for the galaxies).
For the {\sc Hi} field the error power spectrum, $P_{\rm err-HI}$, is a combination of the modeling error estimated from the simulations after fitting the bias parameter, and the noise power spectrum. The noise changes the amplitude of $P_{\rm err-HI}$, especially on small scales, and also introduces an angular dependence. We have indicated this by the $\mu$ dependence in $P_{\rm err-HI}$.  Note the data automatically include shot-noise, since we have a single realization of the halo field in the simulation.  

For the galaxies the error power spectrum, $P_{\rm err-g}$, is due to a combination of shot noise and modeling error estimated from the simulations with fitted bias parameters. For the photometric sample we also need to include the smearing of the density field along the line of sight. 
We include this by damping the signal in the likelihood term with a Gaussian smoothing kernel:
\begin{equation}
    \sum_{\bk, |\bk|\sim k}  |\delta_{\rm g}^b(\bk) - \delta_{\rm g}^{\rm obs}(\bk)| \rightarrow \sum_{\bk, |\bk|\sim k}  |\delta_{\rm g}^b(\bk) - \delta_{\rm g}^{\rm obs}(\bk)|
    \exp\left(-\frac{k^2 \mu^2}{2\sigma_{\rm ph}^2}\right)  \qquad .
\label{eq:photosmoothing}
\end{equation}
Here $\sigma_{\rm ph}$ is the photometric smoothing scale along the line of sight. It is equal to $\simeq 180\,h^{-1}$Mpc at $z=1$ and $\simeq 100\,h^{-1}$Mpc at $z=4$.

\section{Results}

\begin{figure}[h]
    \centering
    \resizebox{0.73\columnwidth}{!}{\includegraphics{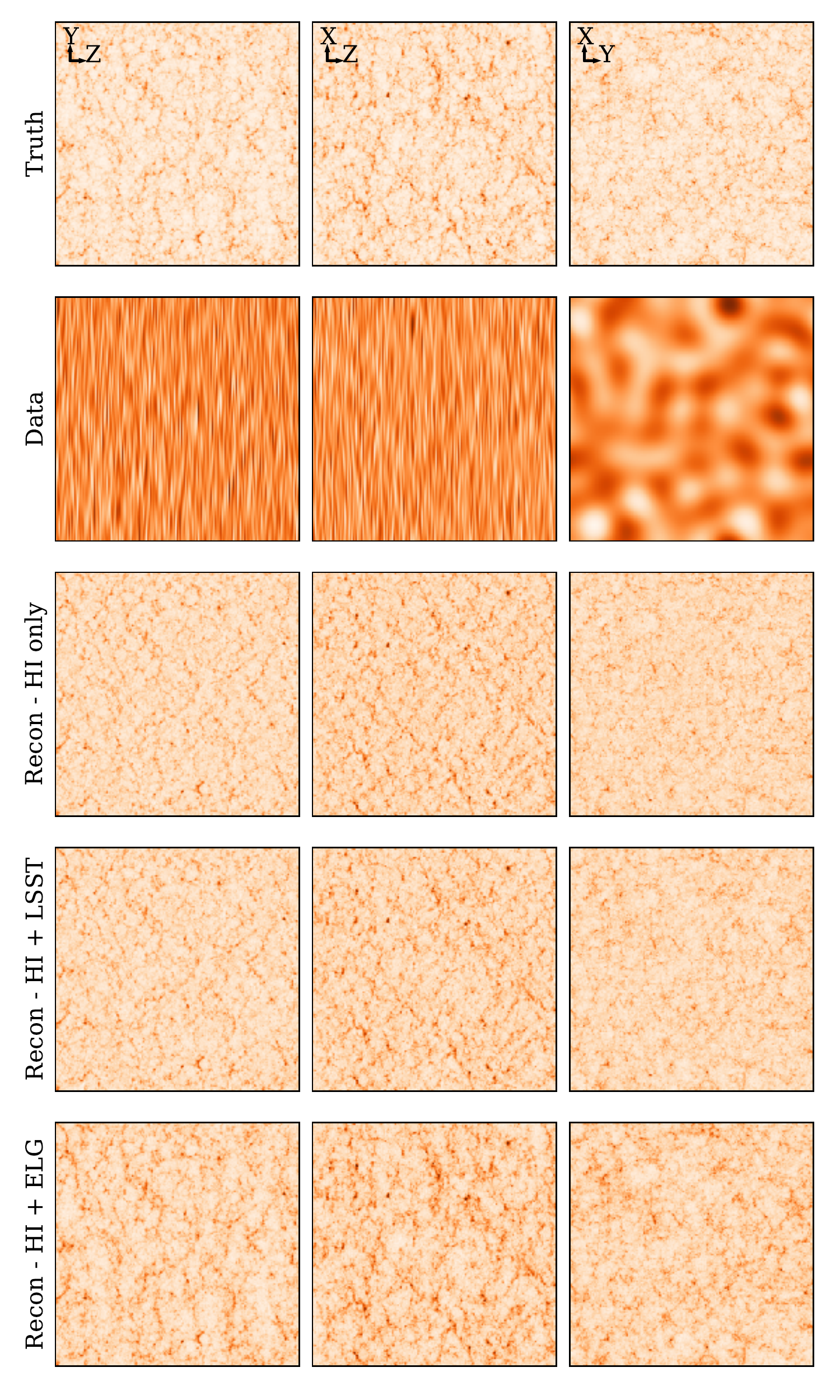}}
    \caption{Slices perpendicular to the 3 box axes of the true H{\sc i} field; the data with thermal noise and wedge; and the reconstructed H{\sc i} field with H{\sc i} data only, H{\sc i}+LSST and H{\sc i}+ELG data at $z=1$. We project over $80\,h^{-1}$Mpc slices, transverse directions show the full box ($1024\,h^{-1}$Mpc). The color scale is the same for all rows and columns except the data row.}
    \label{fig:field}
\end{figure}

In this section we present the results for our reconstructions.  Our primary metrics to gauge the performance of our model and reconstruction are the cross correlation function, $r_{cc}(k)$, and the transfer function, $T_f(k)$, defined as
\begin{equation}
    r_{cc}(k) = \frac{P_{XY}(k)}{\sqrt{P_{X}(k) P_{Y}(k)}} \qquad , \qquad
    T_f(k) = \sqrt{\frac{P_{Y}(k)}{P_{X}(k)}}  \quad ,
\label{eq:rt-def}
\end{equation}
The cross correlation, $r_{cc}$, measures how faithfully the reconstructed map describes the input map, up to rescalings of the output map amplitude. For better visual clarity, we instead show the error power spectrum ($1-r_{cc}^2$) with lower values indicating better reconstruction. The transfer function, on the other hand, tells us about the amplitude of the output map as a function of scale, with $r\,T_f = P_{XY}/P_X$.
These metrics will always be defined between either the model or the reconstructed fields as $Y$, and the corresponding true field as $X$ unless explicitly specified otherwise.

We have to consider 2 sets of bias parameters, one for H{\sc i} and the other for the galaxies. We keep them fixed under the assumption that they have been estimated prior to reconstruction. For the H{\sc i} field at $z=1$, we also smooth the initial density field with $R=6\,h^{-1}$Mpc to estimate quadratic terms as described earlier.
The dynamics in the forward model is taken to be the Zeldovich approximation (ZA). 
All of the data are in redshift space, with the forward model using the ZA velocities to perform the translation.
The reconstruction procedure is outlined in detail in ref.~\cite{Modi19b} but, briefly, it is done in a series of optimization steps. 
We begin on a $256^3$ grid and smooth the likelihood term in Eq. \ref{eq:posterior} for both galaxies and {\sc Hi} on small scales \rsp{i.e.\ multiply the residuals with Gaussian kernel} to fit the large scales first. 
This smoothing is reduced in 5 steps (16, 8, 4, 2, $0\,h^{-1}$Mpc), each with 100 iterations.
More details on this annealing scheme can be found in \cite{Modi18, Modi19a}.
Note that this smoothing is different from the smoothing of the linear field to estimate quadratic components of the bias model discussed earlier and is part of the optimization scheme, not the forward model.
The reconstructed field is then upsampled to $512^3$ grid and smoothing is reduced in 2 steps (2 and $0\,h^{-1}$Mpc) each with 100 iterations.

\begin{figure}[h]
    \centering
    % \resizebox{\columnwidth}{!}{\includegraphics{figs/lsst_N0512.pdf}}
    \resizebox{\columnwidth}{!}{\includegraphics{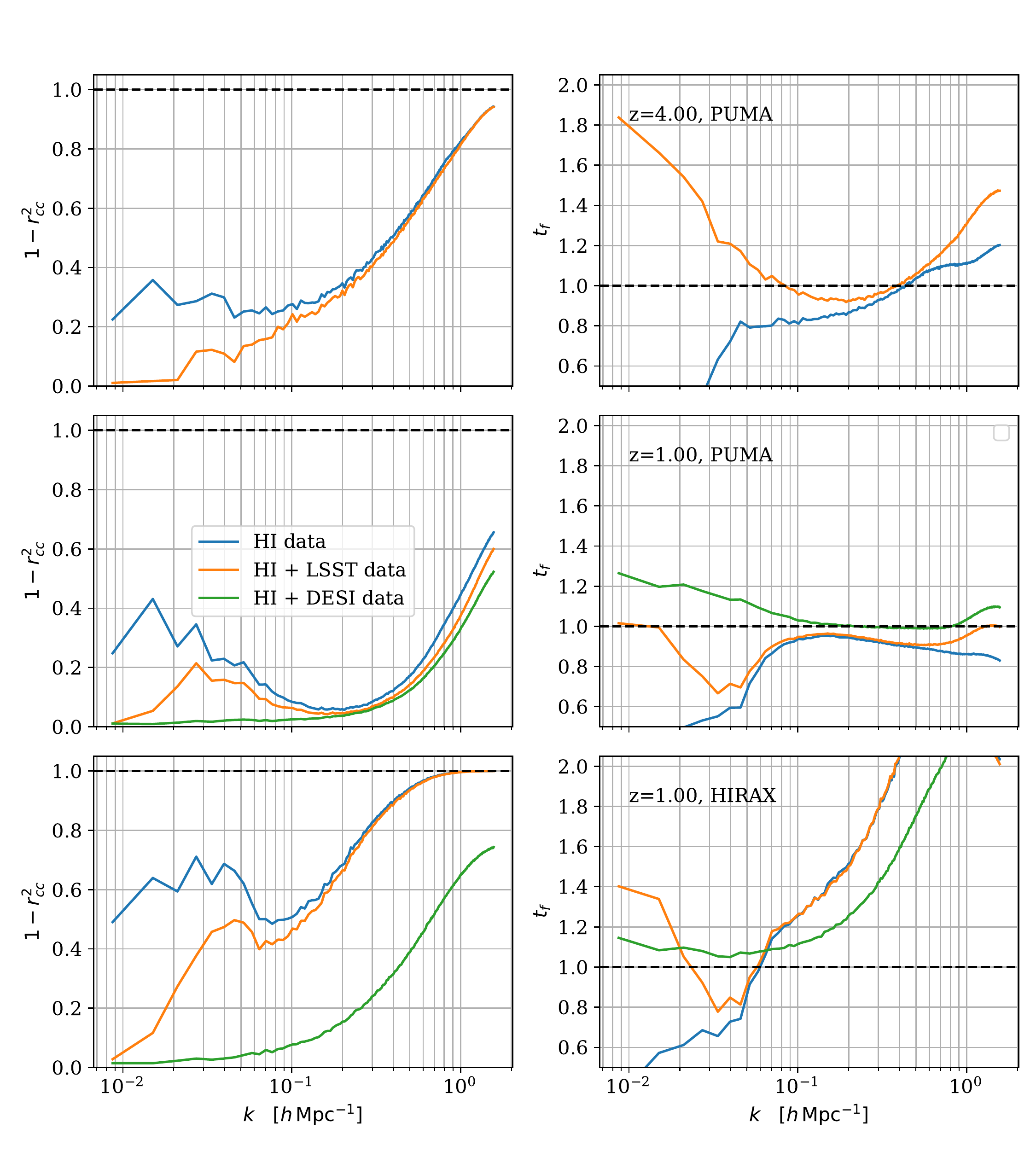}}
    \vspace{-5mm}
    \caption{{Cross correlation (left) and transfer function (right) of the reconstructed H{\sc i} field for PUMA noise level at $z=4$ (top) and $z=1$ (middle), and with HIRAX noise level at $z=1$ (bottom). We show results for reconstruction without galaxies (blue), with a spectroscoic DESI-ELG like sample of $\bar{n} = 10^{-3}\,h^3\,{\rm Mpc}^{-3}$ (green), and a photometric LSST-like sample (orange) of $\bar{n} = 5\times 10^{-2}\,h^3\,{\rm Mpc}^{-3}$ and $\bar{n} = 3.5\times 10^{-3}\,h^3\,{\rm Mpc}^{-3}$ at $z=1$ and $z=4$ respectively. 
    % In dashed and dotted line, we show the statistics for the corresponding reconstructed galaxy sample when applicable.
    }}
    \label{fig:recon}
\end{figure}

\begin{figure}[h]
    \centering
    \resizebox{\columnwidth}{!}{\includegraphics{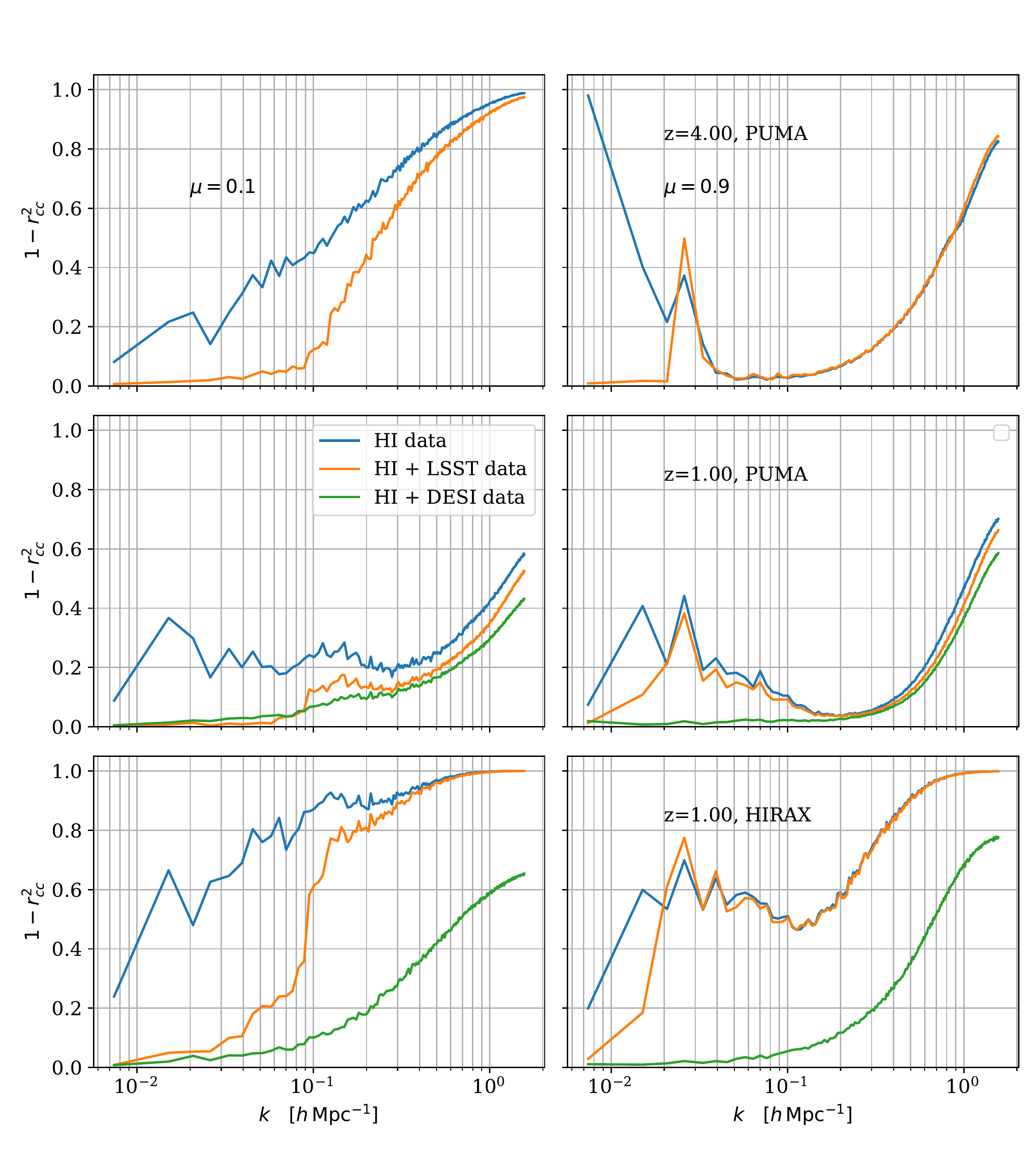}}
    \vspace{-5mm}
    \caption{{Same as Figure \ref{fig:recon} but showing the cross-correlation for the corresponding fields in wedges perpendicular to the line of sight (left, $\mu = [0, 0.2]$) and along the line of sight (right, $\mu \in [0.8, 1]$).
    }}
    \label{fig:recon2D}
\end{figure}

We begin by showing the data and reconstruction at the level of the fields in Figure \ref{fig:field} for $z=1$.
The first two rows show the true H{\sc i} field and the H{\sc i} data with thermal noise and foreground wedge.
The next three rows show the H{\sc i} reconstructed field when the reconstruction is done with only H{\sc i} data, H{\sc i} and LSST data as well as H{\sc i} and ELG data. 
Reconstruction is closest to the underlying truth when we combine H{\sc i} with ELG data, but improvement with LSST data is also apparent in X-Z and Y-Z projections.

Figure \ref{fig:recon} shows the results at the level of the two point function for the following three cases: $z=4$ with PUMA and $z=1$ with both PUMA and HIRAX.
For $z=1$ we show results with both photometric LSST and spectroscopic DESI-ELG data, while at $z=4$ we only have the photometric galaxy sample. 
For comparison, we also show the reconstruction with H{\sc i} data only.
In each case, we see an improvement in the cross-correlation between the reconstructed H{\sc i} field and the true H{\sc i} field as compared to the case when reconstruction is done only with H{\sc i} data.
At $z=1$, we find that the reconstruction with data from a spectroscopic survey far outperforms the reconstruction with a photometric survey, even though the latter has 50 times higher number density.
At $z=4$, as the photometric smoothing scale decreases for LSST, the reconstruction improves and we can recover the largest scales almost perfectly. 
Additionally, in every case we find that more power is reconstructed at the largest scales for H{\sc i} than the original case, as shown in the transfer function. \rsp{Thus we are recovering more structure when we add information from different tracers, however at the same time it consistently biased high.  If uncorrected, this can lead to potential biases. Previous work has suggested correcting for this using simulations \cite{Seljak17, Modi18}, but implementing such a correction is beyond the scope of this work.}

In Figure \ref{fig:recon2D}, we show the same results for cross-correlation but in $\mu$ bins along and perpendicular to the line of sight $\mu \in[0-0.2]$ and $\mu \in[0.8-1.0]$.
We remind the reader that our goal was to supplement the large scale modes lost in the foreground wedge, especially those perpendicular to the line of sight, with modes in galaxy clustering surveys.  Both spectroscopic and photometric surveys probe the perpendicular modes, but the latter loses the line of sight ones.  
Figure \ref{fig:recon2D} clearly shows that the large-scale modes perpendicular to the line of sight are reconstructed very well.
Furthermore, when reconstructed with H{\sc i} data, the LSST field also recovers the modes along the line of sight that are otherwise missing due to photometric uncertainties.
The combination of 21-cm data and LSST is therefore  greater than the sum of its parts.
A translation of these metrics into performance gains for particular science goals can be found in \S 6 of ref.~\cite{Modi19b}.

\FloatBarrier

\begin{figure}[h]
    \centering
    \resizebox{\columnwidth}{!}{\includegraphics{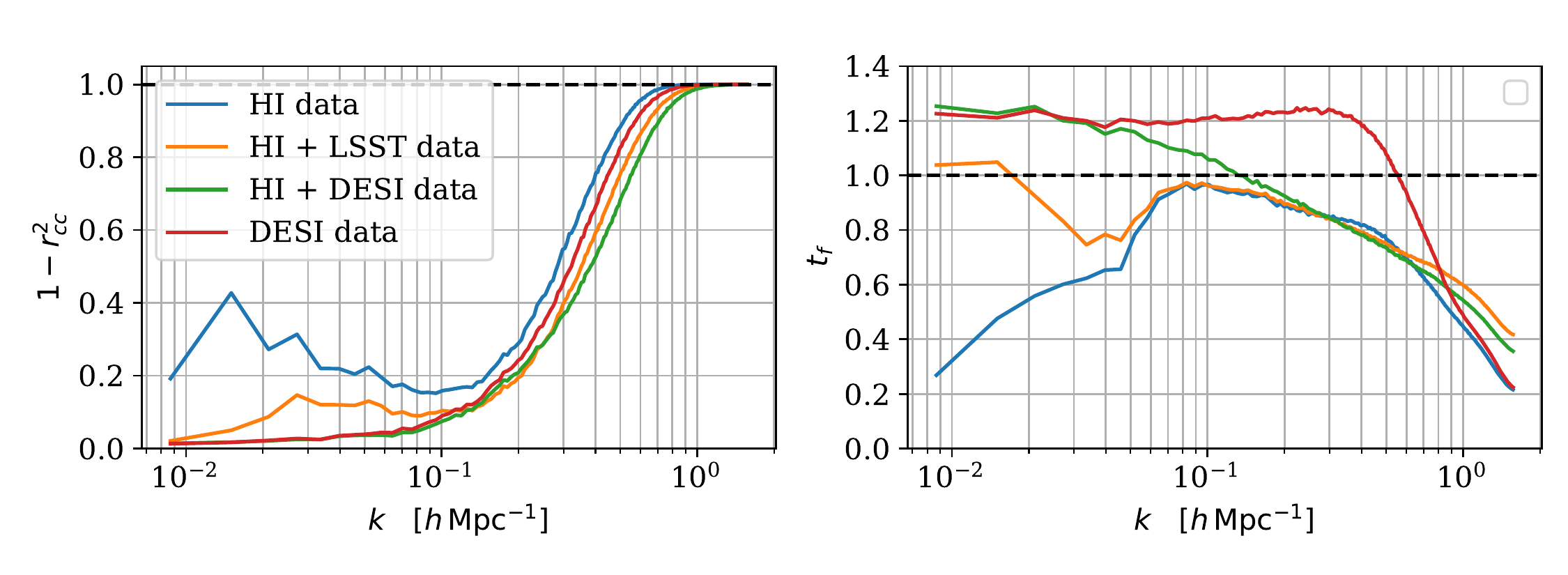}}
    \vspace{-5mm}
    \caption{{Angle average cross correlation and transfer function for the reconstructed initial field with the true initial field at $z=1$ for PUMA noise-levels. We show the results when reconstruction is done with only H{\sc i} data, only DESI ELG-like data, $\bar{n} = 10^{-3}\,h^3\,{\rm Mpc}^{-3}$, the combination of the two and H{\sc i} with LSST data.
    }}
    \label{fig:icrecon}
\end{figure}

In addition to filling in the foreground wedge of H{\sc i} data, we can also use the combination of the different surveys to explore other synergies between the different cosmological tracers.
For instance we can ask how well  our method reconstructs the primordial initial conditions that are shared by, and give rise to, both the observed tracer fields.
In Figure \ref{fig:icrecon}, we show the cross-correlation coefficient and transfer function for reconstructed initial conditions from data at $z=1$ with the true initial conditions.
We compare the cases when reconstruction is done only with a single tracer, such as  H{\sc i} data or ELG-like data with the combination of surveys i.e.\ H{\sc i} and ELG or H{\sc i} and LSST data.
On large scales, DESI and LSST data dominate, improving the reconstruction over using only H{\sc i} field.
However on small scales, H{\sc i} allows us to improve over the galaxy fields and push further into the non-linear regime. 
The combination of the different probes yields higher returns than each tracer considered independently. 

\section{Conclusions}
\label{sec:conclusions}

Interferometric 21-cm surveys have the potential to map out the distribution of matter in the largest cosmological volumes with good radial resolution. However they must contend with bright foregrounds that, when coupled to instrument imperfections, can lead to a loss of large scale modes in the foreground wedge. In this work we build upon the forward modeling approach of Ref.~\cite{Modi19b}, that reconstructed these long-wavelength fluctuations by exploiting the non-linear coupling between small and large scales induced by gravitational evolution, by adding external galaxy distribution data that can help to fill in the missing modes.
Specifically, we consider supplementing the 21-cm data from mock HIRAX and PUMA surveys at different redshifts with a spectroscopic sample (good radial resolution but low number density) and a photometric sample (high number density but poor radial resolution).
The mock galaxies for these datasets are loosely modeled on DESI-ELG sample at $z=1$ and LSST sample at $z=1$ and $z=4$ respectively.

We find that the spectroscopic sample reconstructs the modes significantly better than the photometric sample, despite having much lower number density.
Both galaxy samples are able to reconstruct the largest modes ($k < 0.1\,h\,{\rm Mpc}^{-1}$) transverse to the line of sight very well 
\rsp{with $r_c > 95\%$ for PUMA noise levels at both the redshifts}.
However the contribution of an LSST-like photometric sample to scales smaller than $k\simeq 0.1\,h\,{\rm Mpc}^{-1}$ is not significant,
especially for the thermal noise of HIRAX survey.
At the same time, we find that 21-cm data also reconstructs the correlations in the LSST data along the line of sight on small scales that are otherwise lost due to photometric smoothing. 
The spectroscopic sample, on the other hand, improves reconstruction across all the scales,
especially for a noise-dominated survey like HIRAX where the reconstruction with only 21-cm data is poor \rsp{($r_c=60\%$ even at $k= 1\,h\,{\rm Mpc}^{-1}$)}.
We also explore the synergies of different surveys in reconstructing the initial density field and
find that the combination of surveys performs better \rsp{($r_c = 90\%$ at $k= 0.1\,h\,{\rm Mpc}^{-1}$)}
than using surveys individually \rsp{(best $r_c = 84\%$ for a single spectroscopic survey)}.

With regards to forward modeling approaches, we find that the smoothing scale plays an important role in quadratic bias model when the data itself lacks any direct information on large scales, such as the H{\sc i} field at low redshifts. In this case, using a large smoothing scale to suppress small scales non-linearities when estimating quadratic fields improves reconstruction of large scale modes in H{\sc i} data. 
Interestingly, this does not seem to be the case at higher redshifts. 
The appropriate numerical procedure for the implementation of an effective field theory when modeling large-scale structure at the field level (that would remove the dependence on the smoothing scale) remains an area of active research, and our results show the importance of understanding this issues at a more fundamental level. We plan on pursuing these directions in future work. 

\section*{Acknowledgments} 

M.W.~is supported by the U.S.~Department of Energy and by NSF grant number 1713791.
This research used resources of the National Energy Research Scientific Computing Center (NERSC), a U.S. Department of Energy Office of Science User Facility operated under Contract No. DE-AC02-05CH11231.
This work made extensive use of the NASA Astrophysics Data System and of the {\tt astro-ph} preprint archive at {\tt arXiv.org}.

\appendix

\section{HiddenValley2}
\label{app:HI}

In this work we have used a second run of the Hidden Valley simulations, first reported in ref.~\cite{Modi19a}.  The HiddenValley2 simulation employs the same code, box size, particle loading and initial conditions but has been run to $z=0.5$ with outputs at $z=1.5$, 1.0 and $0.5$.  In addition to lower redshift outputs we have also updated the model used to populate the dark matter halos in the simulation with H{\sc i}, with parameters recalibrated to the wider redshift range and updated measurements of the abundance and clustering of H{\sc i}.

We make use of an $M_{HI}-M_{\rm halo}$ relation in order to populate our dark matter only simulation with H{\sc i}, much as we did in our earlier work \cite{Modi19a}.  We assume the total H{\sc i} mass in a halo of mass $M_h$ is \cite{Padmanabhan17,Castorina17}
\begin{equation}
    M_{HI}(M_h) = A(z)\left(\frac{M_h}{M_{\rm cut}}\right)^\alpha\ \exp\left[-\frac{M_{\rm cut}}{M_h}\right]
\label{eqn:MHI-Mhalo}
\end{equation}
This H{\sc i} mass is split into a central component and a component that moves with the virial velocity of the halo.  Specifically a fraction $f_{\rm cen}$ of the H{\sc i} mass is taken to reside at the halo center and move with the halo center-of-mass velocity.  The remaining $f_{\rm sat}=1-f_{\rm cen}$ of the H{\sc i} mass has an additional, Gaussian line-of-sight velocity distribution with dispersion equal to the virial velocity dispersion of the halo.  For numerical convenience we implemented this by dividing this H{\sc i} into $N_{\rm sat}=\lfloor 1+(0.1M_h/M_{\rm cut})^{0.5}\rfloor$ equal parts and for each drawing an additional line-of-sight velocity component from a Gaussian.  As we found in refs.~\cite{Modi19a,Modi19b} that the details of how we treated such fingers of god were largely unimportant for our science, we have opted to take the simple modeling approach here. Following Figure 7 of ref.~\cite{VN18}, we model the fraction of H{\sc i} in satellites as:
\begin{equation}
    f_{\rm sat} = {\rm min}\left[0.8,\, \frac{0.5 \times (\log M_h - 9.5)^2}{12.8 - 9.5}  \right]
    \quad \forall{M_h > 10^{9.5}\,h^{-1}M_\odot}
\end{equation}
otherwise $f_{\rm sat}=0$.

\begin{figure}
    \centering
    \resizebox{\columnwidth}{!}{\includegraphics{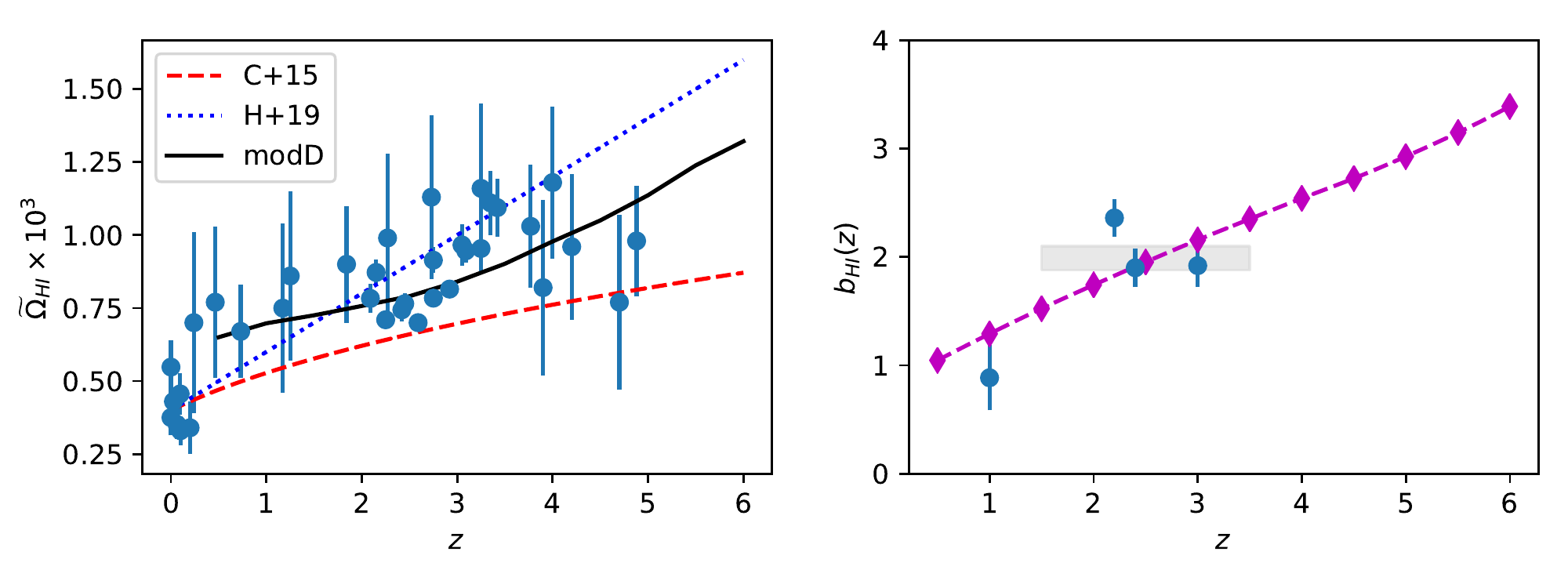}}
    \caption{(Left) The abundance of H{\sc i} as a function of redshift (points; \cite{Zwaan05,Braun12,Martin10,Delhaize13,Rhee13,Lah07,Rao06,Rao17,Noterdaeme12,Songaila10,Zafar13,Crighton15,Bird17,Hu19}).  The dotted line shows a linear fit, $\tilde{\Omega}_{HI}=(0.4+0.2\, z)\times 10^{-3}$, from Fig.~14 of ref.~\cite{Hu19} while the dashed line shows the power-law fit, $\tilde{\Omega}_{HI}=0.4(1+z)^{0.4}$, of ref.~\cite{Crighton15} and the solid black line shows our fiducial model.  (Right) The H{\sc i} bias as a function of redshift.  The error on the $z\approx 1$ point is dominated by the uncertainty in $\tilde{\Omega}_{HI}$.}
    \label{fig:OmegaHI}
\end{figure}

\begin{figure}
    \centering
    \resizebox{0.8\columnwidth}{!}{\includegraphics{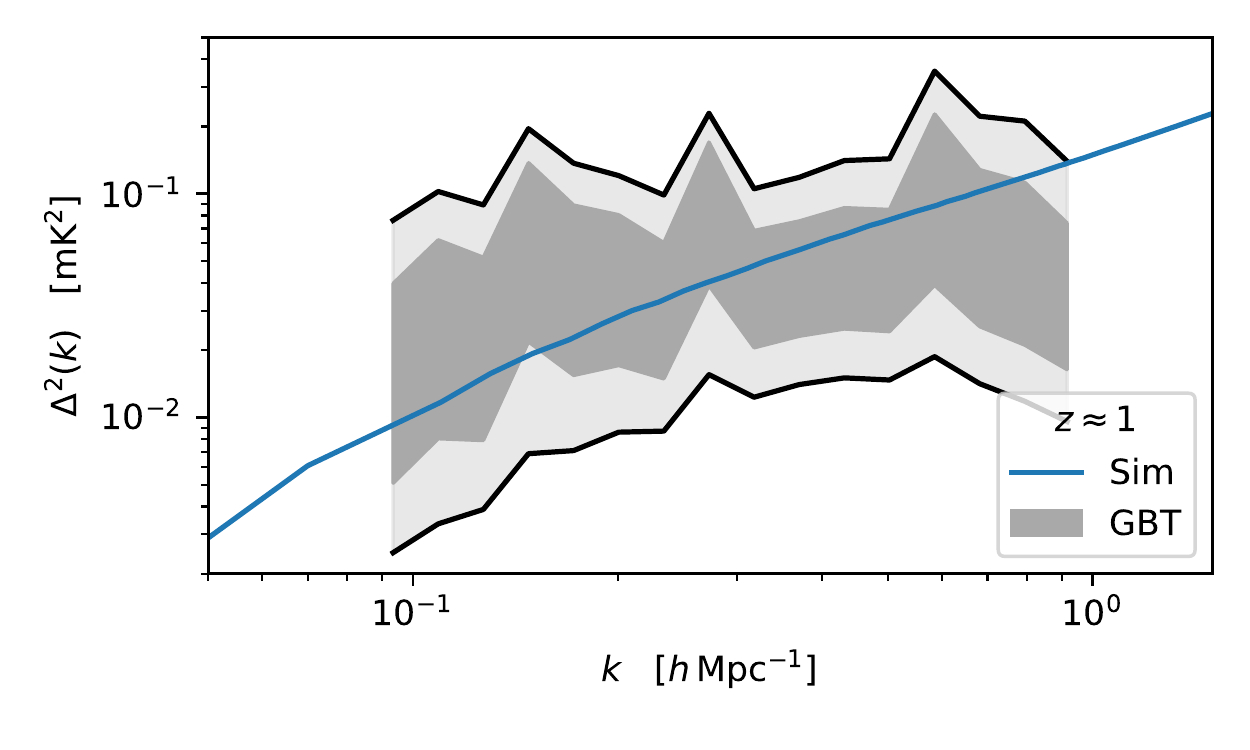}}
    \caption{The monopole redshift-space clustering of H{\sc i} at $z = 1$ from our model, compared to the GBT measurements of refs.~\cite{Masui13,Switzer13} at $z\simeq 0.8$ (grey bands showing 68\% and 95\% confidence regions).  The upper limits from the {\sc Hi} auto power are approximately independent in each bin but the lower limits are perfectly correlated (see ref.~\cite{Switzer13}).
    }
    \label{fig:HIz1}
\end{figure}

There is unfortunately very little data with which to tune the parameters of Eq.~(\ref{eqn:MHI-Mhalo}).  The amplitude, $A(z)$, is largely constrained by the abundance of H{\sc i}, $\tilde{\Omega}_{HI}(z)$ (Fig.~\ref{fig:OmegaHI}).
We have followed the common convention in absorption line studies and H{\sc i} intensity mapping and quoted the abundance as a comoving H{\sc i} density divided by the (physical) $z=0$ critical density.  However we have used a tilde to distinguish this quantity from the more common usage of $\Omega$ as a ratio of (physical or comoving) density at $z$ to critical density at $z$.  The agreement with the data above $z\approx 0$ is quite good.
The values of $\alpha$ and $M_{\rm cut}$ are less constrained.  Both physical intuition and numerical simulations suggest that there is a minimum halo mass ($M_{\rm cut}\sim 10^9-10^{10}M_\odot$) below which neutral hydrogen will not be self-shielded from UV photons.  Above this mass the amount of H{\sc i} should increase as the halo mass increases, though not necessarily linearly (however, simulations suggest $\alpha\approx 1$ at $z>2$).
The characteristic halo mass scale ($M_{\rm cut}$) is determined by the clustering of H{\sc i}.  At $z\simeq0$ an analysis of H{\sc i}-selected galaxies in the Sloan Digital Sky Survey constrains the HOD \cite{Obuljen19}. At $z\simeq 1$ the large-scale bias is known approximately from cross-correlation with optical galaxies \cite{Masui13,Switzer13,Anderson18,Wolz21}, however at higher $z$ there are no direct measurements.  The clustering of Damped Ly$\alpha$ systems (DLAs) has been measured at $z\simeq 2-3$ by ref.~\cite{Rafols18}, and since the DLAs contain the majority of the H{\sc i} at those redshifts this can be used as a proxy for the H{\sc i} clustering amplitude.  The numbers inferred agree reasonably well with an analysis of the most recent hydrodynamical simulations (see Table 6 of ref.~\cite{VN18}).  Based on these considerations we take $\alpha = (1+2z)/(2+2z)$,
\begin{equation}
    A(z) = 1.7 \times 10^9\left( 1+z \right)^{-5/3} \ h^{-1}M_\odot
    \quad , \quad
    M_{\rm cut} = 6\times 10^{10}\exp\left(-\frac{3z}{4}\right) \ h^{-1}M_\odot
    \quad .
\end{equation}

Figure \ref{fig:OmegaHI} shows the abundance and large-scale bias of H{\sc i} predicted from our model (`Model D') compared to observations.  Our model has sufficient flexibility to fit the available data, while also being in broad agreement with the current generation of hydrodynamic simulations.  From Fig.~\ref{fig:OmegaHI} it appears our model overpredicts the clustering at $z\approx 1$, however this is partly due to the way the large-scale bias is estimated in the observations.  We take a closer look at the agreement at $z\simeq 0.8$ (we use the $z=1$ output of our simulation) in Fig.~\ref{fig:HIz1}.  Here we compare the product, $\bar{T}^2\ \Delta^2_0(k)$, predicted by our simulation to the range allowed by the H{\sc i} auto-correlation and {\sc Hi}-WiggleZ cross-correlation \cite{Masui13,Switzer13}.  While there may be some evidence for more small-scale power in the model than the observations, the level of agreement is quite good for most of the range and within the errors on the observation for all scales.

\bibliographystyle{JHEP}
\bibliography{main}
\end{document}